\documentstyle[psfig]{mn}

\title[Hubble Deep Field] {Galaxy morphology to $I=25$~mag in the {\em
  Hubble Deep Field}} \author[Abraham et al.]{R. G. Abraham$^1$,
  N. R. Tanvir$^1$, B. X. Santiago$^1$, R. S. Ellis$^1$, K. Glazebrook$^2$,\\
 {\LARGE\rule{0mm}{6mm}\& S. van den Bergh$^3$}\\ \\
$^1$Institute of Astronomy, University of Cambridge, Madingley Rd.  
Cambridge CB3 OHA\\ 
$^2$Anglo-Australian Observatory, P. O. Box 296, Epping, NSW
2121, Australia\\ 
$^3$Dominion Astrophysical Observatory, National Research Council of Canada, 
5071 W. Saanich Rd. Victoria, B. C. V8X 4M6 Canada} 
\date{Received: Jan. 31, 1996 \ \ \ Accepted: Feb. 8, 1996}
\pagerange{\pageref{firstpage}--\pageref{lastpage}}

\def\etal{{\em et al.}}

\begin{document}

\label{firstpage}

\maketitle

\begin{abstract}
  The morphological properties of galaxies between $21 {\rm~mag} < I <
  25 {\rm~mag}$ in the {\em Hubble Deep Field} are investigated using
  a quantitative classification system based on measurements of the
  central concentration and asymmetry of galaxian light. The class
  distribution of objects in the {\em Hubble Deep Field} is strongly
  skewed towards highly asymmetric objects, relative to distributions
  from both the {\em HST Medium Deep Survey} at $I < 22 {\rm~mag}$ and
  an artificially redshifted sample of local galaxies.  The steeply
  rising number count-magnitude relation for
  irregular/peculiar/merging systems at $I < 22 {~\rm mag}$ reported
  in Glazebrook \etal\ (1995a) continues to at least $I=25~{\rm mag}$.
  Although these peculiar systems are predominantly blue at optical
  wavelengths, a significant fraction also exhibit red $U-B$ colours,
  which may indicate they are at high redshift. Beyond Glazebrook
  \etal 's magnitude limit the spiral counts appear to rise more
  steeply than high-normalization no-evolution predictions, whereas
  those of elliptical/S0 galaxies only slightly exceed such
  predictions and may turn-over beyond $I \sim 24~{\rm mag}$. These
  results are compared with those from previous investigations of
  faint galaxy morphology with HST and the possible implications are
  briefly discussed. The large fraction of peculiar/irregular/merging
  systems in the {\em Hubble Deep Field} suggests that by $I\sim 25
  {\rm~mag}$ the conventional Hubble system no longer provides an
  adequate description of the morphological characteristics of a high
  fraction of field galaxies.

\end{abstract}


\section{Introduction}

The {\em Hubble Deep Field\/} (HDF) is a four square arcminute area of
sky at RA 12h 36m 49.4s~Dec +62$^\circ$ 12$^\prime$
58.0$^{\prime\prime}$ (J2000) observed during December 1995 with the
{\em Hubble Space Telescope} (HST).  The images, obtained during 150
orbits with the Wide Field and Planetary Camera 2 (WFPC-2), represent
the deepest optical imaging survey yet undertaken (Williams
\etal~1996).  Four broadband filters were used whose wavelength
coverage samples the range between the ultraviolet and near infrared
(roughly {\em UBVI}).  In the $I$-band (F814W), the HDF data reaches
$\sim 3$~mag fainter than the deepest ground-based observations at
similar wavelength (Smail {\em et~al.}  1995), and $\sim 1$~mag
fainter than the deepest $I$-band observations undertaken previously
with the HST. The position of the HDF was pre-selected to be a
``typical'' sample of the high galactic latitude sky in terms of
galaxy surface density, while also avoiding bright stars or other
objects whose presence might compromise the depth of the observations.

\begin{figure*}
\centerline{\psfig{figure=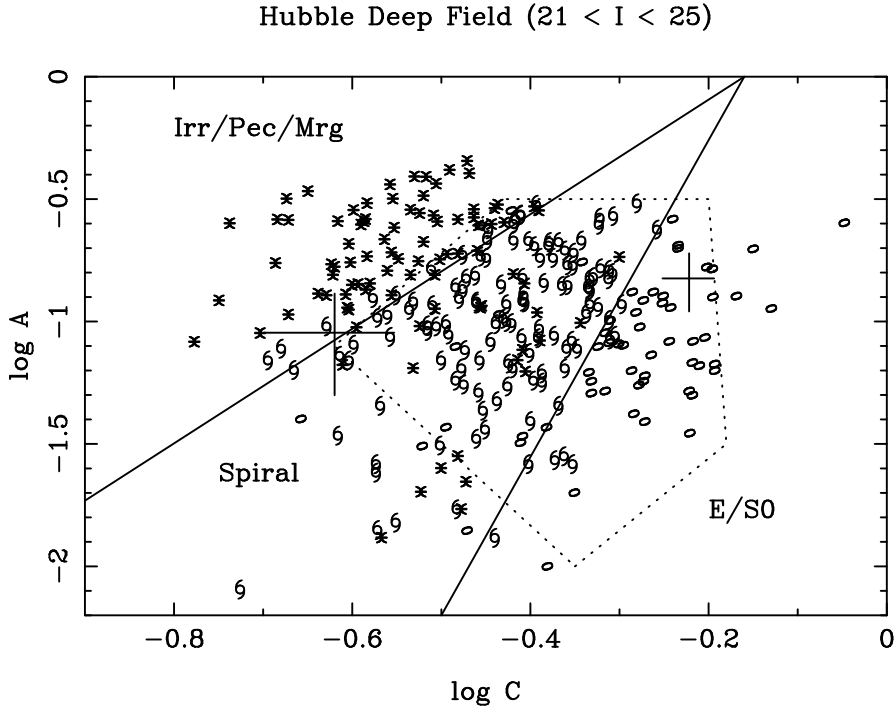,width=5.3in,angle=-90}}
\caption{Asymmetry vs. central concentration for galaxies in the
Hubble Deep Field. RSE's visual morphological classifications are keyed
to the plot symbols: E/S0's are shown as ellipses, spirals earlier
than Sd are shown as spirals, and irregulars/peculiars/mergers are
shown as asterisks.  The sectors subdivide the diagram into regions
where each of these morphological types dominates.  Representative
error bars are shown. The dotted polygon shows the ``convex hull''
enclosing the artificially redshifted local sample of objects
described in the text.}
\end{figure*}

A major objective of the HDF program is to determine the nature of the
rapidly evolving faint blue galaxy population seen in ground-based
spectroscopic and photometric surveys conducted over the past decade
(Broadhurst, Ellis, \& Shanks 1988; Tyson 1988; Colless \etal\ 1990;
Ellis 1990; Koo \& Kron 1992; Lilly 1993; Tresse \etal\ 1993; Cowie
\etal\ 1994; Glazebrook \etal\ 1995b). Many of these objects are
distant ($z > 0.3$), and currently only the HST provides the
resolution needed to characterise their morphological properties.
Results to date have proved difficult to reconcile however, in part
perhaps because of the inherently qualitative nature of galaxy
classification.  Number counts from a sample of 301 galaxies with
\hbox{$I<22$} mag taken from 13 WFPC-2 fields from the {\em Medium
  Deep Survey} (MDS) have been reported by Glazebrook~{\em et al.}
(1995a). These authors classified galaxies by eye and found the
number-magnitude relationships for elliptical and spiral populations
to be consistent with the predictions of high-normalization
no-evolution cosmological models, but that the number of
irregular/peculiar (possibly merging) galaxies were at least an order
of magnitude greater than predicted by the same models.  A similar
conclusion was reached by Driver {\em et~al.} (1995) based on number
counts from a single deeper MDS WFPC-2 field to $I$=24.25~mag and by
Abraham {\em et~al.} (1996) from automated morphological
classification of 521 galaxies from a larger number (21) of WFPC-2 MDS
fields to $I$=22~mag.

In contrast, HST observations reported by Schade {\em et~al.} (1995)
of a small but spectroscopically complete sample of 32 galaxies from the
Canada-France Redshift Survey (CFRS) indicate that the morphological
mix at $z \sim0.75$ is similar to that seen locally, except for a
population of ``blue nucleated galaxies'' (BNGs) which Schade {\em
et~al.\/} suggest comprise $\sim35\%$ of the faint blue galaxy
population. The existence of another major new population of
morphologically distinct objects, linear ``chain galaxies'' that may
be proto-galactic systems, has recently been reported by Cowie {\em
et~al.} (1995) on the basis of HST observations of faint objects
($I>23$~mag) identified in the Hawaii Deep Survey.

A major advantage of the HDF dataset over the earlier images of
Glazebrook \etal, Schade \etal, Driver \etal, and Cowie \etal~ is the
improved signal-to-noise level arising from the considerably longer HDF
exposure time. The exposure times available to those authors were 5-8
hours in F814W, which is substantially less than the $\simeq$35 hours
devoted to the HDF in this filter. The improved signal to noise at
faint limits enables us to extend the work already begun with the MDS,
whose wider spatial coverage nicely complements the HDF data.

Although interesting questions (which the HDF may ultimately answer)
concern the very faint ($I>$25 mag) galaxy counts, in this short note
we are concerned with understanding the morphology and colour
distributions of the $\sim 300$ objects brighter than $I$=25 mag. As
described below, at this limit the bulk of the galaxy population is
detectable in all four HDF passbands, and morphological classifications
are as robust as was the case in our earlier analyses of the MDS data
to $I$=22 mag.

A plan of the paper follows. In \S2 we review our morphological
classification techniques in the context of the HDF data. Section~3
discusses the colour distribution as a function of morphology and, in
\S4, we combine our dataset with that of the MDS, enabling us to
extend Glazebrook \etal~(1995a)'s morphological galaxy count diagram
three magnitudes fainter. A number of new trends appear whose
implications are briefly discussed.

\section{Morphological Analysis of the Hubble Deep Field}

Working with the ``dithered'' (resampled by a factor of 2.5 and
stacked) version 1 HDF images released by STScI, we constructed a
source catalog using the APM faint-object photometry software (Irwin
1985). At very faint ($I \ga 27 {\rm ~mag}$) levels, the number of
objects in the resulting catalog depends sensitively on the smoothing
scale, detection threshold, and minimum object area used by the
software, but in the regime where morphological classification is
reliable ($I < 25 {\rm ~mag}$) the final catalog is robust. 
The photometry was calibrated on the instrumental magnitude
system defined by Holtzmann \etal\ (1995).
Glazebrook \etal (1995a) used the same $I$-band calibration
in their MDS analysis
and thus the results are directly comparable.
The structural properties of galaxies on the
$I$-band frame in the magnitude range $21 {\rm ~mag} < I <25 {\rm
  ~mag}$ were investigated with the automated morphological analysis
system used by Abraham~{\em et al.} (1996) to characterise objects
from the MDS. Visual classifications were also made by two of us (RSE
and vdB) using the classification system discussed by Glazebrook
\etal\ (1995a). The reader is referred to these earlier papers, and to
Abraham~\etal\ (1994), for further details of these procedures beyond
the outline given here.

In the automated classification procedure, the galaxy images were
first isolated from the sky background by extracting contiguous pixels
at least $1\sigma$ above the sky level.  For each galaxy an asymmetry
index, $A$, was measured by rotating and self-subtracting the galaxy
image from itself. A second parameter in the classification procedure,
central concentration $C$, is determined from measurements of the
intensity-weighted second order moments of the galaxy images.
Asymmetric or disturbed galaxies generally have high values of $A$ and
low values of $C$, ordinary spirals have intermediate values of both
parameters, and early-types are distinguished by high central
concentration.

In order to compare directly the measurements of $C$ and $A$ obtained
from ``dithered'' HDF images to those obtained from the MDS, the 23
galaxies in the HDF with $21~{\rm mag}<I<22~{\rm mag}$ (the magnitude
limit of the Medium Deep Survey) were rebinned to the original
sampling of the WFPC-2 camera and degraded by adding Poisson noise to
simulate the appearance of these objects under typical ($\sim$~5000s)
MDS exposures.  A comparison of measurements from the original and
degraded images indicated that the HDF dithering procedure introduces
a small systematic change in the measured asymmetry value of the HDF
data, $\Delta A = 0.053 \pm 0.02$, relative to undithered MDS data.
This small offset was therefore subtracted from the measured asymmetry
values when comparing the HDF and MDS.  It is emphasized that applying
this correction {\em decreases} the the total number of asymmetric
objects in the HDF.  Random errors were determined by Monte Carlo
simulation (c.f. Abraham~{\em et al.} 1994 and 1996) and are no larger
than $\Delta C = \Delta A=0.07$ to $I$=25 mag. Both $C$ and $A$ depend
on the rest frame isophote level at which the measurements are made.
The $1\sigma$ limiting isophote of the HDF data ($\sim 26.5$ mag
arcsec$^{-2}$) is \hbox{1.5 -- 2~mag arcsec$^{-2}$} deeper than the
corresponding limiting isophote for more typical HST data ({\em eg.}
from the MDS), so bright galaxies in the HDF are probed to larger
radii than corresponding objects in the MDS. Strong $(1+z)^4$
cosmological dimming is expected to roughly synchronize the rest-frame
limiting isophotes for the MDS and HDF samples at faint magnitudes,
but a detailed assessment of this effect must await redshift
information (discussed below).

The distribution of HDF galaxies in the $\log(A)$ vs. $\log(C)$ plane
is shown in Figure~1. The visual classifications of RSE are indicated
by the plot symbols on this Figure, and are in excellent agreement
with classifications based on position on the $\log(A)$ vs. $\log(C)$
diagram. Using the visual classifications and the simulations
described below, three sectors were defined on the diagram,
subdividing the galaxy population into irregular/peculiar/mergers,
spirals, and E/S0 systems (the same bins adopted by Glazebrook
\etal~1995a, and Abraham \etal~1996). The HDF asymmetry values are
skewed markedly towards highly asymmetric galaxies relative to the
MDS, as shown in Figure~2.  (Note that Abraham~{\em et al.} 1996 argue
that the MDS data is itself skewed to high asymmetry with respect to
local samples.)  Because the HDF galaxies are presumably at higher
redshifts than their MDS counterparts, it is important to consider
whether this shift may arise from bandshifting or $k$-correction
biases, {\em i.e.} whether the differences are simply a result of our
observing ever further into the rest-frame ultraviolet for more
distant galaxies.

Only limited constraints are currently available for the redshift
distribution of field galaxies to the faint limits explored in this
paper. Using arguments based on the gravitational lensing signal in
well-constrained clusters, Smail {\em et al.}~(1994) and Kneib {\em et
  al}~(1996) claim that most objects will lie in the range
0.5$<z<$2.5. In order to determine the importance of bandshifting
effects on galaxies at these redshifts, $C$ and $A$ were measured for
Frei {\em et al.}~(1995)'s sample of ``generic'' Hubble types between
$T=-5$ (ellipticals) and $T=6$ (Scd galaxies) after artificially
redshifting them to $z=0.5$, $1.0$, $1.5$, and $2.0$. As described in
Abraham~{\em et al.} (1996), $k$-corrections were applied to each
image using the observed colour at each position on the galaxy to
interpolate between a set of template non-evolving SEDs (kindly
supplied by A.  Arag\'on-Salamanca). The artificially redshifted
galaxies occupied the region of the $\log(A)$ vs. $\log(C)$ diagram
within the dotted polygon shown in Figure~1, which therefore maps out
the portion of this diagram which is expected to be occupied by
distant {\em non-evolving} galaxies with Hubble $T$-types between $-5$
and $6$.  It appears that the majority of objects in the
peculiar/irregular/merger sector of Figure~1 correspond to one or more
of the following: (a) to Hubble types later than Sd, (b) to
significantly evolved intermediate Hubble-type spirals, or (c) to a
family of objects outside the Hubble system ({\em i.e.} merging or
peculiar galaxies).

At this point we introduce the visual classifications made for the
sample by RSE and vdB.  
These classifications are in good agreement with classifications from
the $A-C$ scheme on a one-to-one basis and reproduce well the
boundaries of the classes identified from the simulations discussed
above (Figure~1).  Agreement was particularly good for the spheroidal
systems and, where differences between visual and machine
classifications were found, they generally lay at the boundaries of
the 3 categories.  Both vdB and RSE noted that beyond $I \ga 24$ mag
an increasing proportion of faintest spheroidal and spiral galaxies
were less regular than brighter examples, and, in particular, there
was a notable absence of ``grand-design'' spirals.  Asymmetries in
galaxies often resulted from one or more off-centred knots of
activity, suggestive of either increased star-formation or ultraviolet
signal in very distant sources (and emphasizing the importance of
accounting for bandshifting effects).  A montage of typical galaxies
with $I>$23~mag is given in Figure~3 [Plate~1].

Although some of the irregular galaxies in the HDF appear
spectacularly ``chain-like'', the surface density of strictly linear
systems with multiple knots and narrow transverse extents to
$I=25$~mag is $\la 2.5$~arcmin$^{-2}$, {\em i.e.}\/\/ a relatively
small fraction of the total number of irregulars. There are a larger
number of objects which have a bright nucleus and a faint, one-sided
tail, and the properties of these ``tadpole-like'' galaxies (which are
possibly related to the chain-like systems), will be described in a
future work (van~den~Bergh {\em et al.}, in preparation).

\begin{figure}
\centerline{\psfig{figure=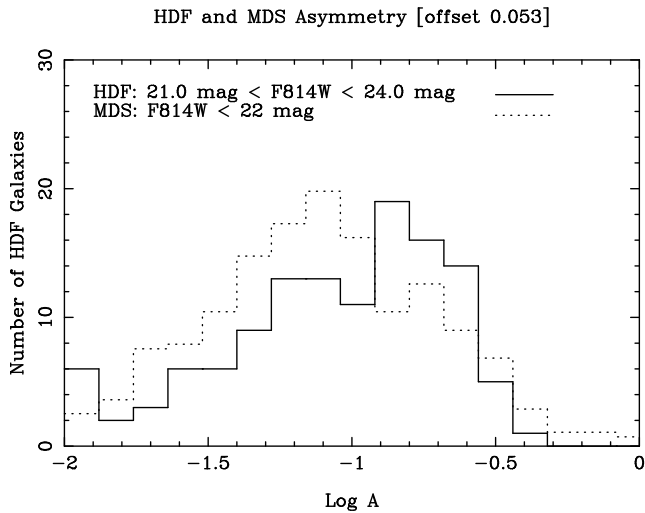,width=3.8in,angle=-90}}
\caption{Histograms of the asymmetry parameter, $A$, for galaxies in
  the Hubble Deep Field (solid line) and Medium Deep Survey (dashed
  line). Note the marked skewness of the HDF data towards asymmetric
  galaxies. The MDS histogram has been normalized to match
  approximately the total counts in the HDF.  The mean asymmetry of
  the HDF histogram has been reduced slightly (as described in the
  text) to correct for the effects of dithering in the images.}
\end{figure}

\begin{figure}
\caption{ (Plate~1). Montage of $I>$23 galaxies from the HDF with: 
  Irr/Pec/Mrg -- columns 1 and 2; Spirals -- columns 3 and 4; and E/S0
  -- columns 5 and 6.  Images were selected on the basis of the
  $A$-$C$ classifier (see Figure~1).}
\end{figure}

\section{Colours}

A very significant advantage of the HDF over earlier data is the
extensive imaging conducted in F300W (hereafter $U$), F450W ($B$) and
F606W ($V$). In the following we will use instrumental magnitudes for
the $UBV$ system since our discussion is only based on relative
colours.  The $I$=25~mag limit for the present investigation not only
corresponds to that point where sound morphological classifications
can be determined (\S2) but also to that limit at which most of the
galaxies are reliably detected in all 4 bands; the $U$ band is the
shallowest exposure and is essentially complete (3 $\sigma$) at $U
\simeq $ 27. Thus, only a few very red objects ($U-I > 2$) in our $I
<$ 25 sample are undetected in U.

Figure~4 [Plate~2] presents the $U-B$ (${\rm F300W}-{\rm F450W}$), $B-V$ (${\rm
F450W}-{\rm F606W}$), $V-I$ (${\rm F606W}-{\rm F814W}$), and $B-I$
(${\rm F450W}-{\rm F814W}$) colours, obtained through
0.4 arcsec apertures, for the HDF sample as a function
of position on the morphological $\log(A)$ vs. $\log(C)$ diagram.  The
colour of an individual galaxy is represented by the colour of the
corresponding symbol on this plot. The trend from one set of passband
colours to the next is remarkably similar. The dominant colour axis
appears horizontal, as expected from the close correspondence between
central concentration and bulge-to-disk ratio (Abraham~{\em et al.}
1994).  The ``faint blue galaxies'' determined from optical colours
correspond to low-central concentration, highly asymmetric objects as
originally claimed by Glazebrook {\em et al.}~(1995a). Objects
classified as early-types are predominantly red, indirectly supporting
the accuracy of classifications made on the basis of $C$ and $A$, and
in agreement with observations suggesting that the early-type
population in the field remains old even at faint flux
levels. 

The distribution of irregular/peculiar/merger objects on the $U-B$
diagram exhibits an interesting behaviour (shown more clearly in
Figure~5). Nineteen of 83 objects in this morphological category
(determined from Figure~1) have {\em blue\/} optical-near infrared
colours ($V-I < 0.6$) and comparatively {\em red\/} UV-optical colours
($U-B > -0.2$).  In fact, objects which are red in $U-B$ tend to have
high asymmetry irrespective of morphological class - a trend which is
not present in other colours.  They are also faint, with only two of
the ``$U-B$ red, $V-I$ blue'' objects being brighter than $I=23$ mag.  One
possibility is that a significant fraction of asymmetric objects have
$z>2.3$ corresponding to the Lyman discontinuity entering the $U$
filter. Alternatively this may be the result of reddening from dust
catalysed by mergers or interactions; this seems less likely as it
would produce a trend in the other colours.
 
\begin{figure}
\caption{ (Plate~2). The $U-B$ (top left), $B-V$ (top right), $V-I$
(bottom left), and $B-I$ (bottom right) colours for the HDF sample as a
function of position on the morphological $\log(A)$ vs. $\log(C)$
diagram. Symbols indicate each object's colour.  Positions on the
morphological diagram are based upon $I$-band morphology. Sectors that
subdivide approximately the ``peculiar/irregular/merger'', ``spiral'',
and ``early-type'' portions of the diagram (defined as for Figure~1)
are shown by solid lines. Note the strong correlation between optical
colour and morphology. Most peculiar systems are optically blue, but
many of these galaxies also exhibit red $U-B$ colours.}
\end{figure}

\begin{figure}
\centerline{\psfig{figure=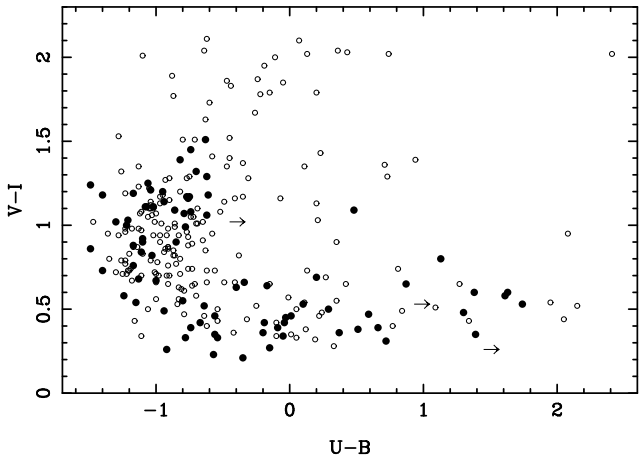,width=3.8in,angle=-90}}
\caption{ The $V-I$ vs $U-B$ colour-colour diagram for all objects with $I<25$ 
  mag. Bold symbols denote objects in the irregular/peculiar/merger
  sector of Figure~1.  The small number of objects with no $U-B$
  colour are represented by lower limits.  Note the plume of objects
  with blue $V-I$ and red $U-B$ colours.}
\end{figure}

\section{Number Counts}

The high signal-to-noise ratio of the HDF data and the reliability of
the morphological classifications allow us to extend the
morphologically segregated galaxy counts 3 magnitudes deeper than was
possible with the Medium Deep Survey. This dramatic improvement
reflects the fact that signal-to-noise is the prime limitation in
image analysis; the increase in angular diameter distance
corresponding to the 3 magnitude improvement is probably very small.
Figure~6 combines the HDF and MDS number counts (with error bars based
on Poisson statistics).  Also shown are the no-evolution predictions
for each type constructed as described in Glazebrook~{\em
  et~al.}~(1995a), adopting Schecheter luminosity functions (LFs) with
parameters given by Loveday {\em et al.} (1992), and a high
normalization ($\phi_\star=0.03 h^3$ Mpc$^{-3}$). The elliptical panel
shows the predicted counts resulting from a flat faint-end slope
($\alpha=-1$), instead of Loveday's original steep faint-end slope.
The very small difference between the counts determined using the
$A-C$ classifications and those determined visually by RSE and vdB are
also indicated in the Figure.

The HDF counts extend the trend already identified by Glazebrook {\em
  et al.} (1995a) to much fainter magnitudes. The over-abundance of
irregular/peculiar/merging systems continues to the $I=25$~mag limit,
at which point these systems represent over 40\% of all galaxies.  As
discussed above, a significant fraction of these systems may lie
beyond a redshift $z\simeq$2.5 and there appears to be a general
evolution of increasing asymetry with redshift. Significantly, some
new trends emerge from Figure~6. Beyond $I=22 {\rm ~mag}$ the spiral
counts now show a significant excess over the no-evolution
predictions. A weaker trend is seen for the spheroidal systems (whose
counts are only marginally above the no-evolution prediction) and
there is some evidence of a turn-over in the last magnitude interval.

The interpretation of Figure~6 depends crucially on both the redshift
distribution of the data and the faint end slope of the luminosity
function. If the local luminosity function is reliably determined from
recent faint surveys (c.f. Ellis {\em et al.} 1996) and the Lyman
limit is always present in distant galaxies, the spiral excess would
appear to imply luminosity evolution for disk galaxies in the interval
1$<z<$2.5. The flattening of the spheroidal galaxy counts beyond
$I=24$~mag in both the visual and automated catalogues is an
interesting, though currently uncertain, result. As this population is
predominantly red, minimal evolution is implied to quite high redshift
and steepening $\alpha$ would make this result yet tighter. The effect
of curvature here is slight: in the absence of luminosity evolution,
the bulk of the $I\simeq$24-25 spheroidals are expected to lie within
$z<$2. In this respect, the decline in the counts is a puzzling result
although it may indicate that the precursors of many present-day
spheroidals lie in one of the other morphological categories.

\begin{figure}
\centerline{\psfig{figure=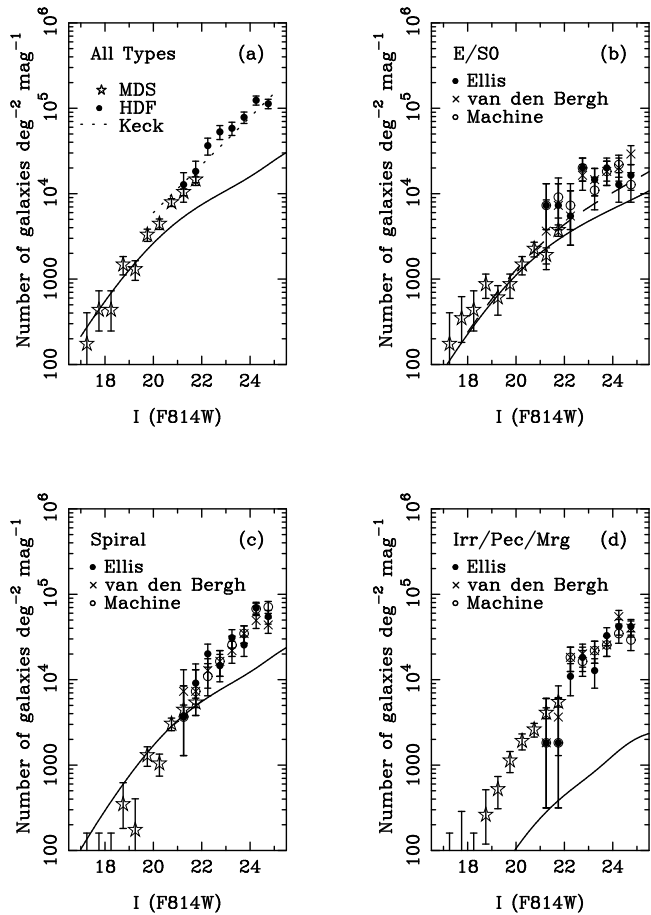,width=3.8in}}
\caption{The number-magnitude relations for morphologically segregated
  samples of galaxies from the HDF and MDS. Open circles indicate
  counts obtained from automated classifications determined from
  Figure~1, closed circles indicate the results from the visual
  classifications of RSE, and crosses indicate the results from the
  visual classifications of vdB. The MDS counts are indicated by the
  stars on each panel. The no-evolution $\Omega=1$ curves from
  Glazebrook~{\em et~al.}~(1995a), extrapolated to $I=25 {\rm ~mag}$,
  are superposed. The dashed line on the E/S0 diagram shows the effect
  of assuming $\Omega=0.1$. The dotted line in panel (a) shows the
  $I$-band number counts determined by Smail~{\em et al.} (1995) from
  two deep fields imaged with the Keck telescope.}
\end{figure}

The large fraction of irregular/peculiar/merger systems detected at
faint magnitudes in the HDF supports earlier suggestions that beyond
$I\sim 24~{\rm mag}$ the conventional Hubble system no longer provides
an adequate description of the structural characteristics for a
significant fraction of galactic (and possibly proto-galactic)
systems.  This has important implications for forthcoming number count
and correlation analyses of the {\em Hubble Deep Field}. The results
from fainter investigations may depend strongly on the prescription
used to define galaxies amongst collections of clumpy
subcomponents. Although the magnitude limit in this paper was chosen
deliberately so as to minimise this ambiguity, the peculiar structure
and obvious distortion of conventional galactic features seen in many
systems at $I<25 {~\rm mag}$ inevitably results in some difficulties
when trying to connect these objects to local counterparts. At $I< 25
{~\rm mag}$ it still seems useful to exploit the (approximate)
connection between sectors on the $A-C$ diagram and conventional
galaxy classifications in order to determine number counts that can be
directly compared with those for the Hubble system. But for fainter
galaxies it may prove more useful to model the distributions of $C$
and $A$ (or other structural parameters) directly, and no longer
require galaxies to be directly mapped onto bins corresponding to
local archetypes.  A quantitative characterisation of the morphologies
of faint galaxies, based on measurements of physically meaningful
structural parameters, provides an objective route forward when the
conventional ``syntax'' of morphological description, based on the
Hubble system and references to local archetypes, has broken down.

\section*{ACKNOWLEDGENTS}

We thank Bob Williams for devoting such a large fraction of Director's
Discretionary Time to the HDF project, and are grateful to the staff
of STScI and ST-ECF for mobilising in order to ensure that the HDF
data was reduced and available quickly. We also thank the UK HST
Support Facility for help in obtaining the HDF data, Daniel Durand of
the Canadian Astronomy Data Centre for help with displaying the images
at DAO, and the members (and former members) of the Cambridge APM
group, especially Steve Maddox and Mike Irwin, for useful discussions
with regard to catalog construction with the APM software.

\end{document}